\documentclass[12pt]{article}
\usepackage{amsmath}
\usepackage{graphicx,psfrag,epsf}
\usepackage{enumerate}
\usepackage{comment}
\usepackage{natbib}
\usepackage{amssymb}
\usepackage[dvipsnames]{xcolor}
\usepackage{bm}
\usepackage{listings}
\usepackage{array}
\usepackage[normalem]{ulem}
\usepackage{multirow}
\usepackage{subcaption}
\usepackage{hyperref}
\usepackage{caption}
\usepackage{booktabs}
\usepackage{longtable}
\usepackage{caption}
\usepackage{setspace}
\usepackage[ruled]{algorithm2e}

\newcommand{\blind}{0}

\begin{document}

\def\spacingset#1{\renewcommand{\baselinestretch}%
{#1}\small\normalsize} \spacingset{1}

\if0\blind
{
  \title{\bf Generalized Conditional Functional Principal Component Analysis}
  \author{Yu Lu$^{1,*}$,
  Xinkai Zhou$^{1}$,
  Erjia Cui$^{2}$,
  Dustin Rogers$^{3}$, \\
   Ciprian M. Crainiceanu$^{1}$, Julia Wrobel$^{4}$, Andrew Leroux$^{3}$ \\
  \small{$^{1}$Department of Biostatistics, Johns Hopkins University} \\
  \small{$^{2}$Division of Biostatistics and Health Data Science, University of Minnesota} \\
  \small{$^{3}$Department of Biostatistics and Informatics, Colorado School of Public Health}\\
  \small{$^{4}$Department of Biostatistics and Bioinformatics, Emory University}}
    \date{}
  \maketitle
} \fi

\if1\blind
{
  \bigskip
  \bigskip
  \bigskip
  \begin{center}
    {\LARGE\bf Generalized Conditional Functional Principal Component Analysis}
\end{center}
  \medskip
} \fi

\bigskip
\begin{abstract}

We propose generalized conditional functional principal components analysis (GC-FPCA) for the joint modeling of the fixed and random effects of non-Gaussian functional outcomes. The method scales up to very large functional data sets by estimating the principal components of the covariance matrix on the linear predictor scale conditional on the fixed effects. This is achieved by combining three modeling innovations: (1) fit local generalized linear mixed models (GLMMs) conditional on covariates in windows along the functional domain; (2) conduct a functional principal component analysis (FPCA) on the person-specific functional effects obtained by assembling the estimated random effects from the local GLMMs; and (3) fit a joint functional mixed effects model conditional on covariates and the estimated principal components from the previous step. GC-FPCA was motivated by modeling the minute-level active/inactive profiles over the day ($1{,}440$ 0/1 measurements per person) for $8{,}700$ study participants in the National Health and Nutrition Examination Survey (NHANES) 2011-2014. We show that state-of-the-art approaches cannot handle data of this size and complexity, while GC-FPCA can.
\end{abstract}

\noindent%
{\it Keywords:}  accelerometry, FoSR, generalized FoSR, generalized FPCA
\vfill

\newpage
\spacingset{1.75} 

\section{Introduction}\label{sec:intro}

Non-Gaussian (binary, counts, categorical, or continuous but skewed and/or heavy tailed) functional data are ubiquitous in real-world applications. Such data can either be collected as non-Gaussian \citep{gaston2008,Kass2001,Kelly2012,sebastian2010,senturk2014,bothwell2022pattern,staicu2012} or transformed into non-Gaussian for interpretability \citep{swihart2015,gaynanova2020,cui2022fui}. For example, the US National Health and Nutrition Examination Survey (NHANES) collected accelerometry data between 2011-2014 and released it in December 2020. Data includes minute-level physical activity summaries for $14{,}693$ participants over seven days, collected using wrist-worn accelerometers. Physical activity at each minute is summarized using a continuous measure called Monitor Independent Movement Summary units (MIMS) \citep{john2019mims}. The distribution of MIMS at each time point can be quite skewed and the interpretation of MIMS units is neither transparent nor easily translatable into actionable information. For instance, a recommendation of "100 MIMS units per day" lacks context, making it difficult to act upon, as the meaning of 100 MIMS units is unclear. A simple way to enhance interpretability is to apply a threshold to the MIMS units, categorizing periods of activity and inactivity. {\it While thresholding inevitably reduces the amount of information available, it provides a simplified and highly interpretable measure that can be easily understood and communicated.} 

For the purposes of this paper we start with the minute level MIMS observed for each individual over multiple days. After excluding $1{,}090$ participants with no "valid days" of wear, $36$ participants who recorded $0$ minutes of activity throughout the day, and $4{,}234$ individuals under age $18$, the remaining data were transformed into minute-level active/inactive indicators. This transformation used a threshold of $10.558$ MIMS per minute for physical inactivity, as proposed  by \citep{karas2022comparison}. This process provides a $1{,}440$ dimensional vector of zeros and ones for each study participant corresponding to each minute of the day. Therefore, the active/inactive profiles in the NHANES data set are stored in a $9{,}333\times 1{,}440$ dimensional matrix, where each row corresponds to a study participant and each column corresponds to a minute of the day.

\begin{figure}[!tbh]
\centering
\includegraphics[width=\textwidth]{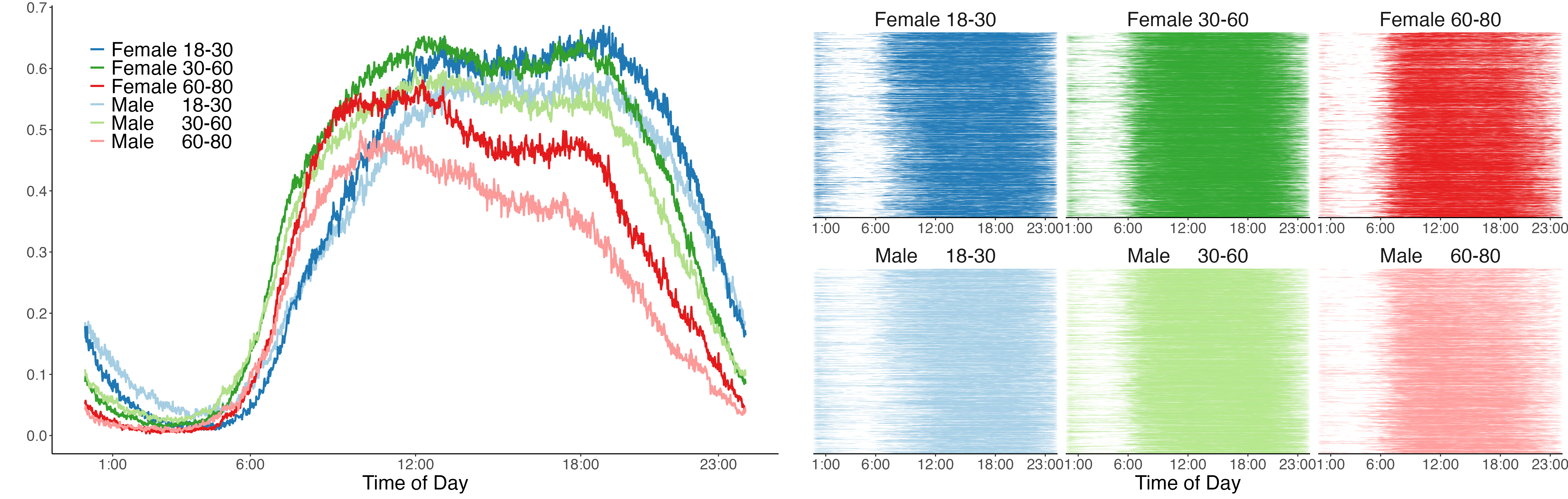}
  \caption{Left panel: empirical probability of being active within each age/gender group. Right panels: individual binary activity profiles for all study participants within each age/gender subgroup. Active minutes are color-coded by the age/gender group (e.g., females between $18$-$30$ are shown in dark blue).}
  \label{fig:fig1}
\end{figure}

The left panel in Figure \ref{fig:fig1} displays the empirical probability of being active across several age/gender groups as a function of time of day. The right panels in Figure \ref{fig:fig1} display the active inactive profiles for all study participants over $18$ in six age/gender groups. The x-axis in all these panels represents time in minutes from midnight to midnight. The lines shown in the left panel are obtained by averaging the binary active/inactive indicators across all participants for each minute; color is used to identify the same age/gender group. For example, the dark blue line in the left panel is the estimated probability of being active during the day for females between $18$ and $30$. This is obtained by averaging by columns the individual profiles shown  in the top left corner of the right panels (titled Female 18-30).   

The plots indicate a strong age effect for both women and men over $30$ (compare the dark green and dark red as well as the light green and light red curves in the left panel). The reduction in physical activity with age is especially pronounced between $8$AM and $12$PM. In contrast, men between $18$ and $30$ have lower activity levels from $8$AM to $12$AM and higher activity levels from $12$AM to $12$PM (compare dark blue to dark green and dark red curves in the left panel). Additionally, within the same each age group, females are more likely to be active than males (comparing lighter and darker shades of the same color). Our goal is to conduct joint inference on the covariates such as gender and age (fixed effects) and physical activity, while accounting for the strong within-person correlation exhibited by the right panels in Figure~\ref{fig:fig1} (random effects). The problem is especially difficult because the functional data (active/inactive indicators at every minute of the day) are not Gaussian.

Several methods have been proposed to model non-Gaussian functional data. For example, the Fast Univariate Inference (FUI) \citep{cui2022fui, sergazinov2023case} focuses on estimating functional fixed effects, but requires multiple observations at each time point, and does not provide decomposition of functional random effects. The Generalized Functional Additive Mixed Models (GFAMM) \citep{scheipl2015famm, Scheipl2016GFADM} implemented in the {\ttfamily refund::pffr()} function \citep{refund} provides a flexible class of models designed for data structures similar to the NHANES. Unfortunately, the methods do not scale up to  the size of NHANES. Generalized Multilevel Function-on-Scalar Regression and Principal Component Analysis (GenMFPCA) \citep{goldsmith2015} is another approach designed for this scenario, but, like GFAMM, it is limited to smaller data sets. The fast Generalized Functional Principal Components Analysis \citep{leroux2023fast}, a binary FPCA method using a fast variational approximation to the Bernoulli likelihood \citep{wrobel2019registration} and Generalized Multilevel Functional Principal Components Analysis (GM-FPCA) \citep{zhou2023gmfpca} provide scalable solutions for conducting FPCA on non-Gaussian data, but do not include fixed-effect inference. A recently proposed method based on Bayesian MCMC \citep{sunkowal2024} could potentially be adapted to the context considered here, but to date it was implemented for Gaussian functional data only. 

To model data of the size and complexity of the non-Gaussian functional data in NHANES, we propose generalized conditional functional principal components analysis (GC-FPCA), which models jointly the fixed and random effects. This is essential for studies such as NHANES, where the active/inactive patterns exhibit substantial heterogeneity and non-stationary within-study participant correlations. Compared to previous approaches, GC-FPCA estimates principal components conditional on the fixed effects. This is achieved by combining three modeling innovations: (1) fitting local generalized linear mixed models (GLMMs) with person-specific random intercept conditional on covariates in windows along the functional domain; (2) conducting a functional principal component analysis (FPCA) on the person-specific functional random effects extracted from the local GLMMs; and (3) fitting a joint GLMM on the complete set of data along the entire functional domain GLMM conditional on covariates and the estimated principal components from the previous step. GC-FPCA allows for flexible modeling of the association between covariates and the functional outcome. While we present GC-FPCA using a varying coefficient model \citep{Hastie1993}, it can be readily extended to  generalized additive models \citep{hastie2017generalized, wood2017gam}, allowing predictors to be smooth, nonlinear functions of covariates.

We will show that GC-FPCA is significantly faster than the state-of-the-art Generalized Functional Additive Mixed Model (GFAMM) \citep{Scheipl2016GFADM}, while achieving similar inferential performance. Unlike GFAMM, GC-FPCA scales up to very large data sets that have become ubiquitous in real life applications. {\it While the application of our method is primarily to binary functional data, methods are general and apply to any other exponential family functional outcomes. In our simulation studies we consider both binary and count data.} 

The rest of the paper is organized as follows.  Section \ref{sec:method} introduces the GC-FPCA approach. Section \ref{sec:simulation} presents a simulation study comparing the computation time and accuracy of our new approach with existing methods. We discuss the NHANES application results in Section \ref{sec:real-data} and conclude with a discussion in Section \ref{sec:discussion}.

\section{Methods}\label{sec:method}
The observed data for participant $i$ is $\{\mathbf{X}_i, Z_i(s), s\in S,\, i = 1,\ldots, I\}$, where $Z_i(s)$ are possibly non-Gaussian measurements, $s\in S$ is a location in the domain $S$ and $\mathbf{X}_i = [X_{i1}, X_{i2}, \ldots, X_{ip}]^T$ is a $p\times 1$ dimensional vector of scalar covariates. 
We assume that $Z_i(s) \sim  {\rm EF}\{\mu_i(s)\}$, where ${\rm EF}(\mu)$ stands for exponential family with mean $\mu$. The GC-FPCA model is
\begin{equation}
\label{eq1}
 g\{\mu_i(s)\}  =  \eta_i(s) = \beta_0(s) + \sum_{r=1}^{p} X_{ir} \beta_r(s) + b_i(s)\; .
\end{equation}
Here $\mu_i(s) = \mathbb{E}\{Z_i(s)\mid \mathbf{X}_i\}$ is the conditional expectation, $g(\cdot)$ is an appropriate link function, $\eta_i(s)$ is the linear predictor, $\beta_0(s)$ and $\beta_r(s)$ are functional fixed effects, and $b_i(s)$ is the subject-specific functional random effect. 

We further assume that $b_i(s) \sim \text{GP}(0, K_b)$ is a zero-mean Gaussian Process with covariance operator $K_b$. 
According to the Kosambi–Karhunen–Loève theorem \citep{karhunen1947}, $b_i(s) = \sum_{l=1}^\infty \xi_{il}\phi_l(s)$, where $\phi_l: S \rightarrow \mathbb{R}$ are orthonormal eigenfunctions such that $\int_S \phi_l^2(s) \, ds = 1$, $\xi_{il} \sim N(0, \lambda_l)$ are mutually independent subject-specific scores across study participants $i$ and eigenvalues $l$, and $\lambda_1 \geq \lambda_2 \geq \ldots$ are the eigenvalues. In practice, we truncate to a finite number of principal components, $L$. With these assumptions model~\eqref{eq1} becomes
\begin{equation}
    g\{ \mu_i(s) \} 
    = \beta_0(s) + \sum_{r=1}^p X_{ir} \beta_r(s) + \sum_{l=1}^L \xi_{il} \phi_l(s)\;.
    \label{eq:GC-FPCA_model}
\end{equation}

\subsection{Fitting local GLMMs with simple random error structure}\label{subsec:binning}
While most methods agree on the general structure of these functional models, the complexity is in how $b_i(s)$ is modeled and estimated. Note that $b_i(s)$ are not directly observed and, in most applications, they capture the remaining subject-specific variation and correlation after accounting for the fixed effects. One of the main innovations of our approach is to observe that models~\eqref{eq1} and \eqref{eq:GC-FPCA_model} can be approximated by local Generalized Linear Mixed Models (GLMM), which can produce initial estimators of $b_i(s)$ at every location $s\in S$ conditional on the fixed effects. This idea is similar to the one used by \citep{leroux2023fast}, who focused on estimating $b_i(s)$ without conditioning on the fixed effects. As conducting inference on the fixed effects is a priority in many applications, it is crucial to understand whether the same idea extends to conditional estimation of the functional random effects.

Specifically, data are divided into local bins along the functional domain. Let $\{s_k:k=1,\ldots,K\}$ be the equally spaced centers of these bins, though the methodology can also be applied to unequally spaced bin centers. Given a bin width $w$, the bin $S_k$ centered at $s_k$ contains the observations at $\{s_{k-\lceil \frac w2 \rceil}, ..., s_k, ..., s_{k+\lceil \frac w2 \rceil}\}$, where $\lceil{x}\rceil$ is the smallest integer greater than or equal to $x$.  The data in bin $k$ is $\{Z_i(s_j), k\}, 1 \leq i \leq I, s_j \in S_k, 1 \leq k \leq K $. For cyclic data, bins near the boundary of the domain may cross over this boundary.

For each of the $K$ bins, $S_1, \ldots, S_K$, we estimate the GLMM 
\begin{equation}
\label{eq:localGLMM}
g[E\{Z_i(s_j \mid s_j \in \mathcal{S}_k)\} ] = \beta_0^*(s_{k}) + \sum_{r = 1}^p X_{ir} \beta_r^*(s_k) + b_i^*(s_{k})\;,\end{equation}
where $b_i^*(s_{k})\sim N(0,\sigma^2_k)$ is a subject-specific random intercept at the bin center $s_k$. Note that in these local models the ``within-subject replicates" are the $\{j:s_j\in S_k\}$ and the random intercept $b_i^*(s_k)$ is assumed to be constant in these models. In the original model~\eqref{eq1} there is no assumption that $b_i(s)$ is locally constant. However, the hope is that if the bins are small enough then the local model in ~\eqref{eq:localGLMM} provides a good approximation for model~\eqref{eq1}. The main goal of this paper is to show that this approximation does, indeed, produce valid estimates of fixed effects in model ~\eqref{eq1}, and provides a computationally scalable alternative to existing methods.  

There are several appealing characteristics of model~\eqref{eq:localGLMM}: (1) it is a simple GLMM with $p+1$ main fixed effects and one random intercept; (2) fitting these local models can be parallelized, though this is often not necessary;  (3) it is straightforward to expand the model to include more complex fixed and random effects structures; (4) partial and complete missing data at the participant level are easily handled when data are missing at random; (5) it can be implemented using a variety of frequentist and Bayesian software; and (6) by borrowing information across participants, including the fitting of the local models, the method can easily handle lack of variation at the study participant level (e.g., if all data are $0$ or $1$ for several study participants).

\subsection{PCA in the linear predictor space}\label{subsec:GFPCA}
The estimated subject-specific random effects, $\widehat{b}_{i}^*(s_{k})$, from model~\eqref{eq:localGLMM}, form a function over $s_k$, $k=1,\ldots,K$. To estimate the covariance of these random effects, we employ the fast covariance estimation (FACE) method \citep{xiao2016}, implemented in the {\ttfamily refund::fpca.face()} function in {\ttfamily R} \citep{R}. This approach provides smooth estimates of the eigenfunctions $\{\hat{\phi}_l(s)\} | l = 1, \ldots, L\}$ of the covariance operator $\widehat{K}_b$, and the number of eigenfunctions, $L$, can be selected using the percent variance explained (e.g., $95$\%, $99$\%). These eigenfunctions are evaluated along a B-spline basis, enabling estimation of the eigenfunctions continuously over the domain $s \in S$, not only at the discrete points $s_k$. The functional parameters for the fixed effects are modeled as $\beta_r(s) = \sum_{m=1}^{M} \beta_{rm} B_m(s)$, $r = 0, 1, \cdots, p$, where $B_1(s), \cdots, B_{M}(s)$ is a set of $M$ basis functions. Here we used the same number of basis function for each fixed effects, but this restriction is not necessary. Model~\eqref{eq1} is then approximated by the following generalized linear mixed effect model (GLMM) with $(p+1)M$ fixed effect slopes and $L$ uncorrelated random slopes. 
\begin{equation}
                g\{ \mu_i(s) \}  = \sum_{m=1}^{M} B_m(s)\beta_{0m}  + \sum_{r=1}^p  \sum_{m=1}^{M}  \{X_{ir}B_m(s)\}\beta_{rm} + \sum_{l=1}^{L} \xi_{il}\widehat{\phi}_l(s)\;,
            \label{eq:full_conditional}
\end{equation}
where $\xi_{il} \sim N(0,\sigma_l^2)$ are mutually independent. While this is still a relatively complex model, it is much more tractable than the initial conceptual model~\eqref{eq1}. 

\subsection{Algorithm and Computional considerations}\label{subsec:algo}

The algorithm for fitting the GC-FPCA is summarized below. 

\begin{algorithm}[H]
  1. Bin data into local windows $S_1, \cdots, S_K$.\\ 
  2. Fit a local GLMM $g[E\{Z_i(s_j \mid s_j\in \mathcal{S}_k)\} ] = \beta_0^*(s_{k}) + \sum_{r = 1}^p x_{ir} \beta_r(s)^* + b_{i}^*(s_{k})$ in each bin $S_k$ and obtain the estimated  random effects $\widehat{b}_{i}^*(s_k)$. \\
  3. Apply FACE to the covariance matrix $\widehat{K}_b(u,v)=\text{Cov}\{\widehat{b}_i^*(u), \widehat{b}_i^*(v)\}$ and obtain the eigenfunction estimates $\widehat{\phi}_l(s)$ for $l = 1, ..., L\}$. \\
  4. Estimate the GLMM 
  $g\{ \mu_i(s) \}  = \sum_{m=1}^{M} \beta_{0m} B_m(s) + \sum_{r = 1}^p X_{ir} \sum_{m=1}^{M} \beta_{rm} B_m(s)  + \sum_{l=1}^{L} \xi_{il}\widehat{\phi}_l(s)$, where $\xi_{il}\sim N(0,\lambda_l)$ are mutually independent.
  \caption{GC-FPCA}
  \label{algo:GFAM}
\end{algorithm}

The computation time for step 1 is negligible. Step 2 requires the fit of $K$ GLMM with one random intercept in overlapping or non-overlapping bins. Assuming a window size $w$ for each bin, the fixed effects design matrix has dimensions $(wI) \times (p+1)$, and the random effects design matrix has dimensions $(wI) \times I$. The computing complexity for each local GLMM is $O(wI)$, and without parallelization, the total computing time for step 2 is at most $O(wIK)$. However, this procedure is easily parallelizable using computer clusters, reducing the computing time to $O(wI)$. Step 3 is also highly optimized, as the computational complexity of FACE is $O(IKc)$, where $c$ is the number of B-spline basis functions used for smoothing, which is typically much smaller than $I$ and $K$. In general, the computational complexity of  steps 1, 2, and 3 is small relative to the final step.

The design matrix for the fixed effects both for GC-FPCA and GFAMM is $(IK) \times [(p+1)M]$ dimensional, where $M$ is the number of basis functions for modeling the fixed effects coefficients. However, the design matrix for random effects for the GC-FPCA method is $IK \times IL$, relative to $IK \times IM_{\rm GFAMM}$ for GFAMM, where $M_{\rm GFAMM}$ is the number of functional effects used to capture the subject-level variability. Thus, the computational complexity of GC-FPCA is reduced by at least a factor of $\left(\frac{M_{\text{GFAMM}}}{L}\right)^3$ compared to GFAMM (complexity of Cholesky decomposition in penalized regression). For example, if GFAMM uses $30$ B-spline basis functions to model models $b_i(s)$, and GC-FPCA uses $L=4$ principal components, computing time is reduced by a factor of at least $500$. The required memory for the design matrix is also reduced by a factor of $\left(\frac{M_{\text{GFAMM}}}{L}\right)$ compared to GFAMM. Moreover, the penalty matrix of the random effect for GC-FPCA is an identity matrix, while GFAMM requires a penalty matrix of dimension $IM_{\text{GFAMM}} \times IM_{\text{GFAMM}}$. Practically, the memory required to fit GFAMM explodes with the number of study participants; for example, attempting to fit GFAMM to a subset of the NHANES data ($N=1000$) required more than 100 Gb of RAM.

\section{Simulation}\label{sec:simulation}

We evaluate the empirical performance of the GC-FPCA and compare it to the GFAMM via a comprehensive simulation study. Functional responses were simulated on an equally spaced grid of $K$ points in $ S \in [0, 1] $. For simplicity, the number of fixed effects was set to $ p = 1 $. For subject $ i $ at time $s_k$, the linear predictor data-generating model was:
\begin{align*}
    \eta_i(s_k) = \beta_0(s_k) + \beta_1(s_k) x_i + \sum_{l=1}^{4} \xi_{il} \phi_l(s_k)\;
\end{align*}

Here, $ \beta_0(s_k) $ is the intercept at time $ s_k $, $ \beta_1(s_k) $ is the fixed effect coefficient for $ x_i $ at time $ s_k $, $ \xi_{il} \sim N(0, \lambda_l) $, $ \{s_k = \frac{k}{K}: k = 0, 1, ..., K\} $, and $ K $ is the number of observations per study participant. We set the true eigenvalues $ \lambda_l = 0.5^{l-1}, l = 1, 2, 3, 4 $. 

We simulated Gaussian, binary and Poisson functional data. Binary functional data are generated as $Z_i(s_k) \sim \text{Bernoulli}\{\mu_i(s_k)\}$, where ${\rm logit}\{\mu_i(s_k)\}=\eta_i(s_k)$. Poisson data are generated as $Z_i(s_k) \sim \text{Poisson}\{\mu_i(s_k)\}$, where $\log\{\mu_i(s_k)\}=\eta_i(s_k)$. Here we present the results for binary data, while the results for Gaussian and Poisson data are provided in the supplementary materials. The functional fixed effects coefficients $\beta_0(s_k)$ and $\beta_1(s_k)$ are generated as: $\beta_0(s_k) = \sum_{m=1}^{14} \beta_{0m} B_m(s_k)$, $\beta_1(s_k) = \sum_{m=1}^{14} \beta_{1m} B_m(s_k)$, where $B_m(s)$ is the B-spline basis matrix with $14$ bases created using knots at $\{0, 0.1, \dots, 1\}$ and $\beta_{0m}$, $\beta_{1m}$, for $m=1,\ldots,M=14$, are the true fixed effect coefficients. 

Data were simulated with different number of subjects, $I = 100, 200, 500, 1000$, and number of observations per subject, $K = 100, 200, 500, 1000$. All simulations used $L = 4$ principal components using either: (1) the Fourier basis $\{\sqrt{2}\sin(2\pi s)$, $\sqrt{2}\cos(2\pi s)$, $\sqrt{2}\sin(4\pi s)$, $\sqrt{2}\cos(4\pi s)\}$, or (2) the orthogonal polynomials basis $ \{1, \sqrt{3}(2s-1), \sqrt{5}(6s^2-6s + 1), \sqrt{7}(20s^3-30s^2+12s-1)\}$. The fixed effect variable, $x_i$, was simulated from a Bernoulli distribution (binary predictors) or a Gaussian distribution (continuous predictors). For each scenario, a total of $1{,}000$ simulations were conducted. Conducting simulations for all combinations of parameters would have been impossible even for our extensive computing resources. Instead, we varied one parameter at a time while keeping the other parameters fixed. For example, to evaluate the effect of different sample sizes, we use Fourier basis functions and binary fixed effects, and set number of observations per study participant $K = 100$ and the percentage of data used for constructing local bins $w = 5\% $ of the number of observations per study participant. The combinations of all these choices resulted in $64$ simulation scenarios. For model evaluation we used different bin lengths corresponding to  $w=2\%, 5\%, 10\%, 15\%, 20\%$ of the number of observations per study participant for step 2, respectively (results shown in the supplementary materials).

\subsection{Model Evaluation Criteria}\label{subsec:eval}

\begin{sloppy}
Methods were compared to GFAMM \citep{scheipl2015famm} implemented in the {\ttfamily refund::pffr()} function. We used $30$ cubic regression basis functions for the fixed and random effects. Comparisons were conducted in terms of: (1) accuracy in estimating fixed effects; (2) inference on fixed effects; and (3) computational efficiency.
    
\end{sloppy}

The accuracy of the linear predictor $\eta(\cdot)$ estimator was assessed using the mean integrated squared error $ \text{MISE}(\eta) = \frac{1}{N} \sum_{i = 1}^N \int_0^1 \{\hat{\eta}_i(s) - \eta_i(s)\}^2 ds$, where $\hat{\eta}_i(s)$ is the linear predictor for study participant $i$ at location $s\in S$. The accuracy of the estimators of functional coefficients $\beta_0(\cdot)$ and $\beta_1(\cdot)$ was assessed using the integrated squared error (ISE) defined as $\text{ISE}(\beta_i) = \int_0^1 \{\widehat{\beta_r}(s) - \beta_r(s)\}^2 ds $ for $r = 0, 1$. The accuracy of the principal components was assessed using the mean ISE over all four eigenfunctions, $\text{MISE}(\phi) = \frac{1}{4} \sum_{l=1}^4 \int_0^1 \{\hat{\phi}_l(s)-\phi_l(s) \}^2 ds $.

The inferential performance of the confidence intervals was evaluated through the empirical coverage probability of $95\%$ pointwise confidence intervals at each location, averaged along the functional domain. This average coverage (AC) across the functional domain, which is defined as $\text{AC}(\beta_r) = \frac{1}{K} \sum_{k = 1}^K \left[ \frac{1}{N} \sum_{n = 1}^{N} \mathbb{I} \{\widehat{\beta}_r(s_k) \in \text{CI}_{rn}(s_k) \} \right],$ where $r = 0, 1$, $\text{CI}_{rn}(s)$ is the confidence interval for $\beta_r(s)$ in iteration $n$, and $N$ is the total number of iterations. These pointwise confidence intervals are constructed using Wald's method, based on linear mixed-effects modeling for the full conditional model in \eqref{eq:full_conditional}.

Given the large number of simulation scenarios, simulation results are shown for binary functional data with fixed number of subjects, $I=100$, and a varying number of observations per function, $K$, as well as for a fixed number of observations per function, $K=100$, and a varying number of subjects, $I$. Results for all other simulation scenarios are shown in the supplementary materials using the same organization and notation. Our method is not sensitive to different bin width used for step 2, a bin width around $10\%$ of the number of total sampling points gives more accurate estimators (on linear predictors, fixed effect coefficients and eigenfunctions) while maintain a speedy computing time. The summary of these simulations results is that: (1) GC-FPCA works well in a variety of scenarios and it is scalable to a large number of study participants; (2) GC-FPCA provides similar or slightly better results than GFAMM, when GFAMM works; (3) GC-FPCA works well in many scenarios when GFAMM does not, especially when the number of study participants increases above $I=500$; and (4) GC-FPCA can be applied to the NHANES active/inactive profiles, whereas GFAMM cannot. 

\subsection{Simulation results: computation times, linear predictors, and fixed effects} 
Tables \ref{table:accuracy_I} and \ref{table:accuracy_K} summarize the simulation results for binary functional data. The bin length was set $5\%$ of the number of observations per study participant and the Fourier basis was used. A median of $52.6\%$ of the functional responses are $1$s for the setting with $I = 100$ and $K = 100$, similar to the proportion of ones for other settings.

Table \ref{table:accuracy_I} presents the simulation results when varying the number of study participants, $I$, with a fixed number of observations per function, $K=100$. The computing time for one data set on a personal laptop ($36$GB RAM, $14$ core GPU) was much shorter for GC-FPCA compared to GFAMM (e.g., $0.06$ minutes compared to $80$ minutes for $I=100$) and on joint high performance computing exchange (JHPCE) ($0.2$ minutes compared to $660$ minutes for $I=200$). For $I=500$ and $I=1{,}000$, GFAMM ran for more than $24$ hours and the program was stopped, while GC-FPCA  ran in $2$ and $15$ minutes, respectively. These differences are partially explained by the discrepancy in the number of elementary calculations, but they accelerate at large sample sizes. This is likely due to the compounding effects of the memory allocation. These computing times are specific to the laptop they were run on and they could vary with configuration and implementation, but they provide a realistic comparisons of what to expect in practice. 
\begin{table}[!htbp] \centering 
\footnotesize
\begin{tabular}{lcrrrrrr}
\toprule

I & Method & Time (min) & ${\rm MISE}(\eta)\times 10$  & ${\rm ISE}(\beta_0)\times 10^2$ & ${\rm AC}(\beta_0)$ & ${\rm ISE}(\beta_1)\times 10^2$ & ${\rm AC}(\beta_1)$\\
\midrule
\multirow{2}{*}{100} & GC-FPCA & 0.06 & 3.31 & 5.79 & 0.91 & 11.48 & 0.91 \\
& GFAMM & 79.83 & 3.88 & 5.46 & 0.92 & 11.34 & 0.91 \\
\midrule
\multirow{2}{*}{200} & GC-FPCA & 0.18 & 2.80 & 2.97 & 0.93 & 5.77 & 0.93 \\
& GFAMM & 656.37 & 3.79 & 2.86 & 0.92 & 6.18 & 0.91 \\
\midrule
\multirow{2}{*}{500} & GC-FPCA & 2.05 & 2.50 & 1.30 & 0.93 & 2.47 & 0.93 \\
& GFAMM & - & - & - & - & - & - \\
\midrule
\multirow{2}{*}{1000} & GC-FPCA & 15.30 & 2.40 & 0.74 & 0.92 & 1.36 & 0.93 \\
& GFAMM & - & - & - & - & - & - \\
\bottomrule
\end{tabular}
\caption{Simulation results for different $ I $ when $ K=100 $. The computation time (\textit{``Time(min
)"}), MISE of $\eta$ (MISE($\eta$)), ISE of $\beta_{p}(s), p=0,1$ (``ISE($\beta_{r}$), $r=0,1$"), and AC of $\beta_{p}(s), p=0,1$ (``AC($\beta_{p}$), $p=0,1$"), reported in the table, are median values across 1000 replications. ``-" stands for computing times exceed 24 hours.}

\label{table:accuracy_I}
\end{table}

\begin{figure}[!tbh]
\centering
\includegraphics[width=\textwidth]{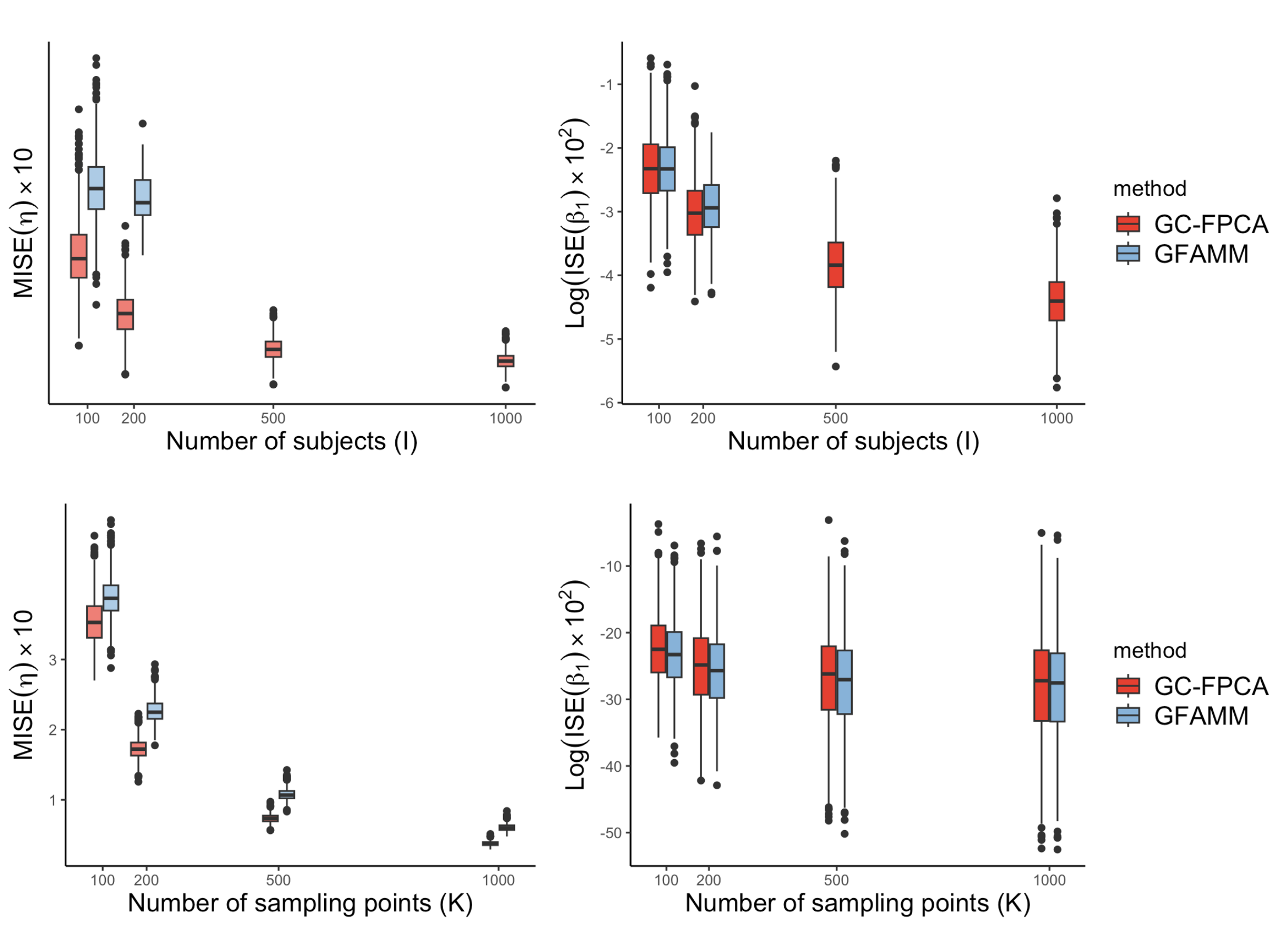}
  \caption{Comparison of simulation results between GC-FPCA (red) and GFAMM (blue) for binary functional data generated with Fourier eigenfunction. First row: varying the number of subjects, $I$, for a fixed number of observations per function, $K=100$. Second row: varying the number of observations per study participant, $K$, for a given number of subjects, $I=100$. First column: MISE for the linear predictor, $\eta$. Second column: Log ISE for the $\beta_1(\cdot)$ fixed effects coefficient function. Third column: Log computation time. All distributions are obtained from $1{,}000$ simulations.} 
  \label{fig:simu-box}
\end{figure}

For a smaller number of study participants ($I=100$ and $I=200$) the estimation accuracy for GC-FPCA is comparable, though GC-FPCA tends to perform slightly better. For example, when $I=200$ the median MISE ($\times 10$) for the linear predictor ($\eta$) is $2.80$ for GC-FPCA and $3.79$ for GFAMM, while the median ISE ($\times 10^2$) for the $\beta_0(\cdot)$ function is $0.93$ for GC-FPCA and $0.92$ for GFAMM. All these differences are very small on the variability scale of these estimators (as will be shown in Figure~\ref{fig:simu-box}). When sample size increases ($I\geq 500$), GFAMM becomes impractical (takes more than $24$ hours per data set). GC-FPCA is still feasible and the estimation accuracy continues to improve for $\beta_0(\cdot)$ and $\beta_1(\cdot)$, though for the linear predictor $\eta_i(\cdot)$ the improvements are smaller. This is not surprising, as adding study participants improves estimation of the population level (fixed effects) parameters, $\beta_0(\cdot)$ and $\beta_1(\cdot)$. However, the number of observations per study participant is kept constant (in this case, $K=100$), and the estimators of the subject-specific linear predictor get better only through the better estimation of the fixed effects.
\begin{table}[!htbp] \centering 
\footnotesize
\begin{tabular}{lcrrrrrr}
\toprule

K & Method & Time (min) & $MISE(\eta)\times 10$  & $ISE(\beta_0)\times 10^2$ & $AC(\beta_0)$ & $ISE(\beta_1)\times 10^2$ & $AC(\beta_1)$\\ \midrule

\multirow{2}{*}{200} & GC-FPCA & 0.06& 1.60 &  4.87 & 0.93 & 9.54 & 0.93 \\

& GFAMM & 61.54 & 2.27 & 4.58 & 0.92 & 9.31 & 0.93  \\
\midrule

\multirow{2}{*}{500} & GC-FPCA & 0.06 & 0.68 & 4.27 & 0.94 & 8.53 & 0.94 \\

& GFAMM & 52.18 & 1.08 & 4.11 & 0.91 & 8.33 & 0.93\\
\midrule

\multirow{2}{*}{1000} & GC-FPCA & 0.13 & 0.18 & 3.85 & 0.94 & 7.81 & 0.94 \\

& GFAMM & 51.53 & 0.33 & 3.83 & 0.90 & 7.77 & 0.93 \\
\bottomrule
\end{tabular}
\caption{Simulation results for different $ K $ when $ I=100 $. The computation time (``Time(min)"), MISE of $\eta$ (``MISE($\eta$)"), ISE of $\beta_{r}(s), r=0,1$ (``ISE($\beta_{p}$), $p=0,1$"), and AC of $\beta_{p}(s), p=0,1$ (``AC($\beta_{p}$), $p=0,1$") reported in the table, are median values across 1000 replications.}
\label{table:accuracy_K}
\end{table}
The top left panel in Figure~\ref{fig:simu-box} displays the distributions of $ 10\times \text{MISE}(\eta) $ of GC-FPCA and GFAMM. This plot supplements the summaries in Table~\ref{table:accuracy_I}, which provides only the median of the MISE values, and supports the results that, in general, GC-FPCA performs at least as well as GFAMM in terms of estimating the linear predictors in this particular scenario. The top right panel in Figure \ref{fig:simu-box} displays the distributions of $ \log\{10^2\times \text{ISE}(\beta_1)\}$ of GC-FPCA and GFAMM, indicating that the performance of the two methods is comparable in terms of estimation of the fixed effects, as well. 

Table~\ref{table:accuracy_K} is similar to Table~\ref{table:accuracy_I}, but summarizes the simulation results for binary functional data for a fixed number of study participants, $I=100$, varying instead the number of observations per function, $K$. In this case, the speed of GC-FPCA was largely unaffected by the increase in the number of observations per study participant, $K$. The reason is that the main computational bottleneck for GC-FPCA is step 4 and not fitting a larger number of local GLMMs in step 2. In this case GFAMM was reasonably fast even in cases when the number of observations per function increased to $K=1{,}000$, though still $500$ times slower than GC-FPCA . The reason for this is likely the fact that GFAMM is highly efficient at smoothing within a person because it uses low-dimensional splines. However, when the number of study participants increases, the design matrix of random effects becomes very large, which limits the scalability of GFAMM in terms of number of study participants. The rest of Table~\ref{table:accuracy_K} shows a similar message with that of Table~\ref{table:accuracy_I}: when GFAMM works, its linear predictor and fixed effects estimation performance is comparable to that of GC-FPCA. The bottom row of Figure~\ref{fig:simu-box} further supports the conclusion that GC-FPCA and GFAMM perform similarly for fixed effects and linear predictor estimation when GFAMM is computationally feasible.

\subsection{Simulation results: principal components}\label{subsec:sim_PC}
\begin{figure}[!tbh]
\centering
\includegraphics[width=0.8\textwidth]{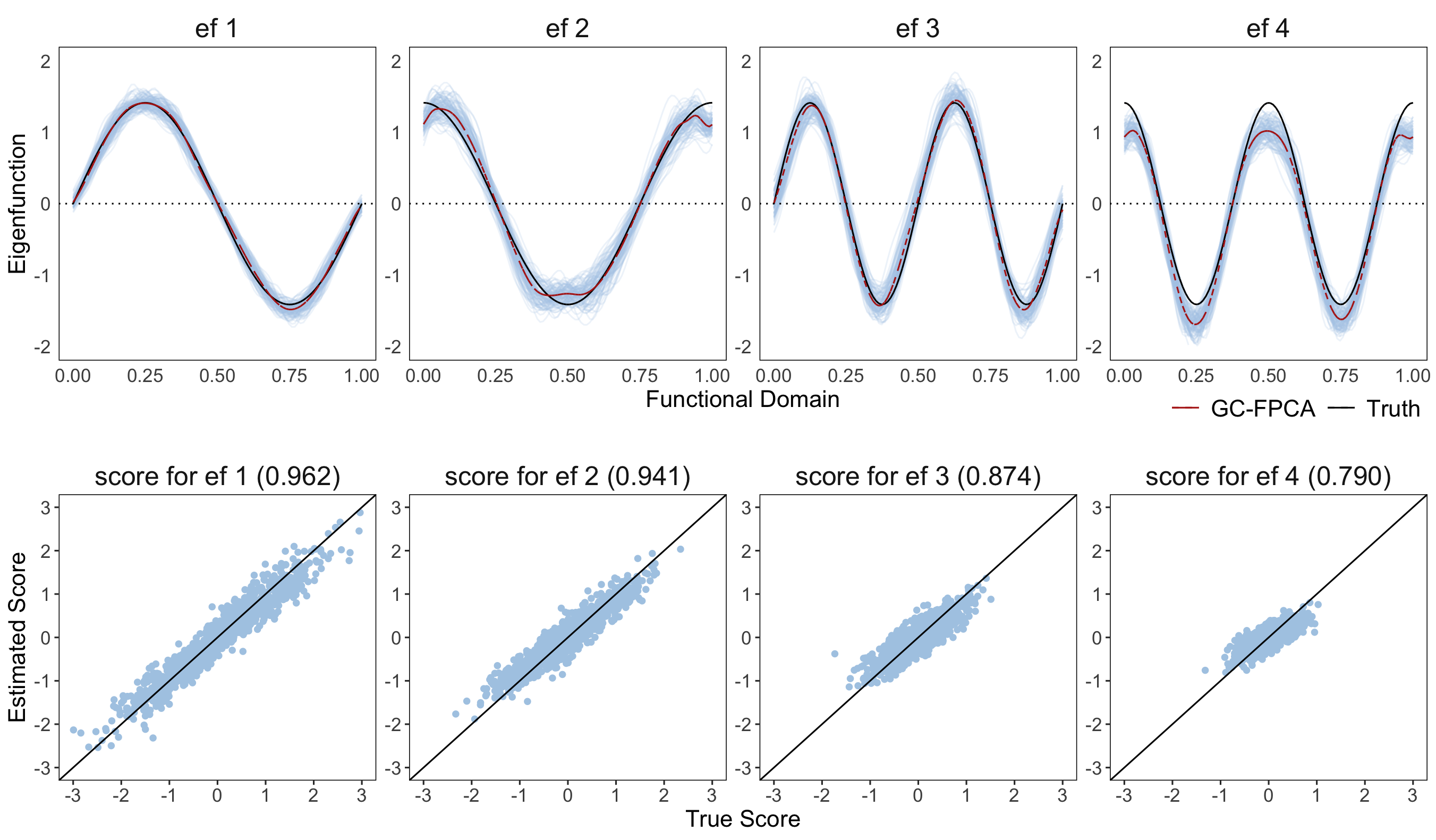}
  \caption{Simulation results of 1000 simulations when $ I = 1000, K = 100$ (binary functional data using Fourier eigenfunctions). First row of panels: $100$ estimated eigenfunctions (red), true eigenfunction (black solid lines), average over $1{,}000$ GC-FPCA estimates of the eigenfunctions (dashed, red line). Second row of panels: scatter plots of true (x-axis) versus estimated (y-axis) scores (blue dots) with their correlation (in the parentheses), identity line (black solid line) for one simulated data set.} 
  \label{fig:simu-ef}
\end{figure}
Figure~\ref{fig:simu-ef} displays the estimated eigenfunctions (first row of panels) and scores (second row of panels) for the simulation setting $I = 1{,}000$ and $K= 100$. The top panels display $100$ estimates of each of the first four eigenfunctions using GC-FPCA as blue solid lines. The black solid curves correspond to the true eigenfunctions and the dark red dashed curves indicate the average of the eigenfunction estimators based on all $1{,}000$ simulations. These results indicate that the estimated eigenfunctions capture the overall shape of the true eigenfunctions. The second row of panels of Figure \ref{fig:simu-ef} displays the scatter plots of the true (x-axis) versus the estimated (y-axis) scores from GC-FPCA in one simulation. These results are representative across simulations. The true and estimated scores exhibit very high correlation ($0.962$, $0.941$, $0.874$, $0.790$ for the first $4$ eigenfunctions, respectively) and follow very closely the identity line (solid black curve). 

Table~\ref{table:simu_ef} provides the median MISE $\times 10$ across $1{,}000$ simulations for more combinations of number of study participants, $I$, and observations per study participant, $K$. Results for all other simulations scenarios are provided in the supplementary material, though the main messages are the same: the MISE for eigenfunctions decreases decreases with the increase in $I$ and $K$.

\begin{table}[ht]
\begin{center}

\begin{tabular}{ccc|ccc}
\toprule
I & K & ${\rm MISE}(\phi) \times 10$ & I & K &  ${\rm MISE}(\phi)\times 10$\\ \midrule
 100 & 100& 2.04 & 100 & 100 &  2.04\\
 200&100& 1.08 & 100&200 & 0.97 \\
 500&100& 0.51 & 100&500 & 0.60\\
 1000&100& 0.33 & 100&1000 & 0.44\\

\bottomrule
\end{tabular}

\end{center}
\caption{Median MISE ($\times 10$) across $1{,}000$ simulations  for estimating the eigenfunctions for different $I$ and $K$ (binary functional data using Fourier eigenfunctions).}
\label{table:simu_ef}
\end{table}

\section{Application to NHANES accelerometry data}\label{sec:real-data}

\subsection{Data description}\label{subsec:data_description}

In Section~\ref{sec:intro} we outlined key characteristics of the US National Health and Nutrition Examination Survey (NHANES), a large-scale, ongoing study led by the United States Centers for Disease Control and Prevention (CDC). NHANES provides a nationally representative sample of the non-institutionalized US population. NHANES collects data in two-year waves, with wearable accelerometers deployed in the 2003-2004, 2005-2006 \citep{troiano2008}, 2011-2012, and 2013-2014 waves \citep{belcher2021,john2019mims}. For this study, we used objectively measured physical activity data obtained from wrist-worn accelerometers from the 2011-2014 waves (released in December 2020). Data were collected at $80$Hz resolution along three axes, then  summarized as Monitor Independent Movement Summary (MIMS) units at the minute level for up to seven days for each study participant.

Out of the $14{,}693$ study participants of the NHANES analytic sample (see description of the NHANES 2011-2014 accelerometry analytic sample in \citep{belcher2021}), $1,090$ participants were excluded for having no "valid days" of wearing and $36$ were excluded for having $0$ minutes of activity throughout the day. For the purpose of our study, we excluded $4{,}234$ study participants under age $18$, and $633$ study participants with missing data on mortality status or covariates. Thus, the sample analyzed in this paper contained a total of $8{,}700$ study participants.

The binary active/inactive profiles were obtained at the minute level by thresholding the daily MIMS data at $10.558$ \citep{karas2022comparison}. More precisely, $Y^B_{ih}(s) = 1\{Y_{ih}(s) \geq 10.558\}$, where $Y_{ih}(s)$ corresponds to the $i^{\text{th}}$ individual's MIMS unit on day $h$ at minute $s$. With up to seven days of accelerometry data per subject, the active/inactive functions were defined as $Z_i(s) = \text{median}\{Y^B_{ih}(s): h = 1,\ldots, M_i\}$, where $M_i$ is the number of days with good quality of data (as described by NHANES) for the study participant $i$. For example, if $M_i=7$, $Z_i(s)$ is $0$ if the study participant was inactive at time $s$ for at least four days, and $1$ otherwise. If $M_i$ is even and there are at least $\frac{M_i}{2}$ active days for time $s$, we define $Z_i(s) = 1$, marking the profile as active.

One could argue that thresholding loses information and complicates the simpler problem of analyzing continuous MIMS values into binary active/inactive profiles. However, we do not argue to not conduct the analysis with the original, continuous MIMS measurements. Instead we argue to conduct complementary analyses on the thresholded data, which provides active/inactive profiles. The reason is interpretation and translation of findings into actionable information. As we pointed out in the introduction, {\it one cannot promote an increase of ``100 MIMS units per day", because no one knows what that is, but one can easily understand and communicate a $10$-minute change from inactive to active minutes per day).}  

\subsection{Analysis results}\label{subsec:NHANES_analysis}
As described in Section~\ref{subsec:data_description}, we start with the participant-specific active/inactive pattern, $Z_i(s)$, and study the time-varying effects of age and gender, while accounting for the sizeable residual variation and correlation. To do that, we use a particular case of model~\eqref{eq1} applied to the probability of being active $p_i(s)=P\{Z_i(s)=1\}$
\begin{eqnarray}
\label{eq:NHANES_model}
{\rm logit}\{p_i(s)\} = \beta_0(s) + \beta_1(s) \text{Age}_i + \beta_2(s) \text{Gender}_i + \sum_{l=1}^{4} \xi_{il}\phi_l(s)\;
\end{eqnarray}
Here, $\beta_r(s) = \sum_{m=1}^{20} \beta_{rm} B_m(s)$, $r = 0, 1, 2$. The term ``gender" is used for consistency with the NHANES 2011-2014 nomenclature as a binary variable (female=1, male=0); this nomenclature may change in future studies. We use $p=2$ fixed effects and $L=4$ principal components, which captured $83$\% of the residual variability on the linear predictor space scale. Fixed effects are modeled using cubic regression splines with $20$ knots. 

\begin{figure}[!tbh]
\centering
\includegraphics[width=0.85\textwidth]{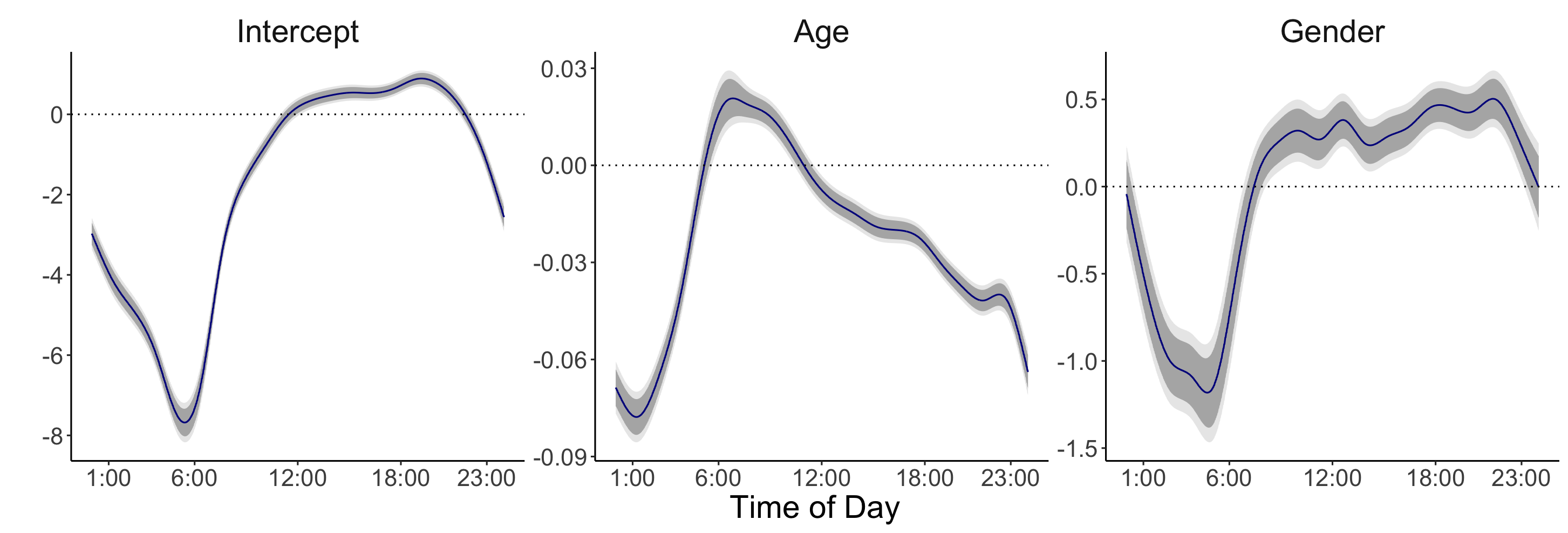}
  \caption{Estimated fixed effects (blue solid lines) in the model~\eqref{eq:NHANES_model} applied to the NHANES data. Pointwise (dark gray) and correlation and multiplicity adjusted (CMA, light gray) 95\% confidence intervals. Panels from left to right: intercept, age, gender (female).}
  \label{fig:nhanes-cov}
\end{figure}

Figure~\ref{fig:nhanes-cov} displays the estimated coefficients with the $95\%$ pointwise (dark gray shaded area) confidence bands and multiplicity adjusted (CMA) confidence intervals (light gray shaded area) confidence bands \citep{crainiceanu2024book}. The left panel corresponds to the estimated functional intercept, $\widehat{\beta}_0(s)$, indicating that, conditional on age and gender, the probability of being active is much lower during the night, increases sharply in early morning, stays high during the day, and then starts decreasing in late afternoon and evening. The middle panel corresponds to the effect of age, which is statistically significant throughout the day, with the exception of two short periods around 6AM and 11:30AM. Results indicate that, conditional on gender, older individuals are less active in the evening, slightly more active in the morning, and far less active in the afternoon. The right panel corresponds to the effects of gender, which indicates that females are less  active during the night, but tend to be more active during the day (recall that $1$ indicates female in this data set). 

\begin{figure}[!tbh]
\centering
\includegraphics[width=0.85\textwidth]{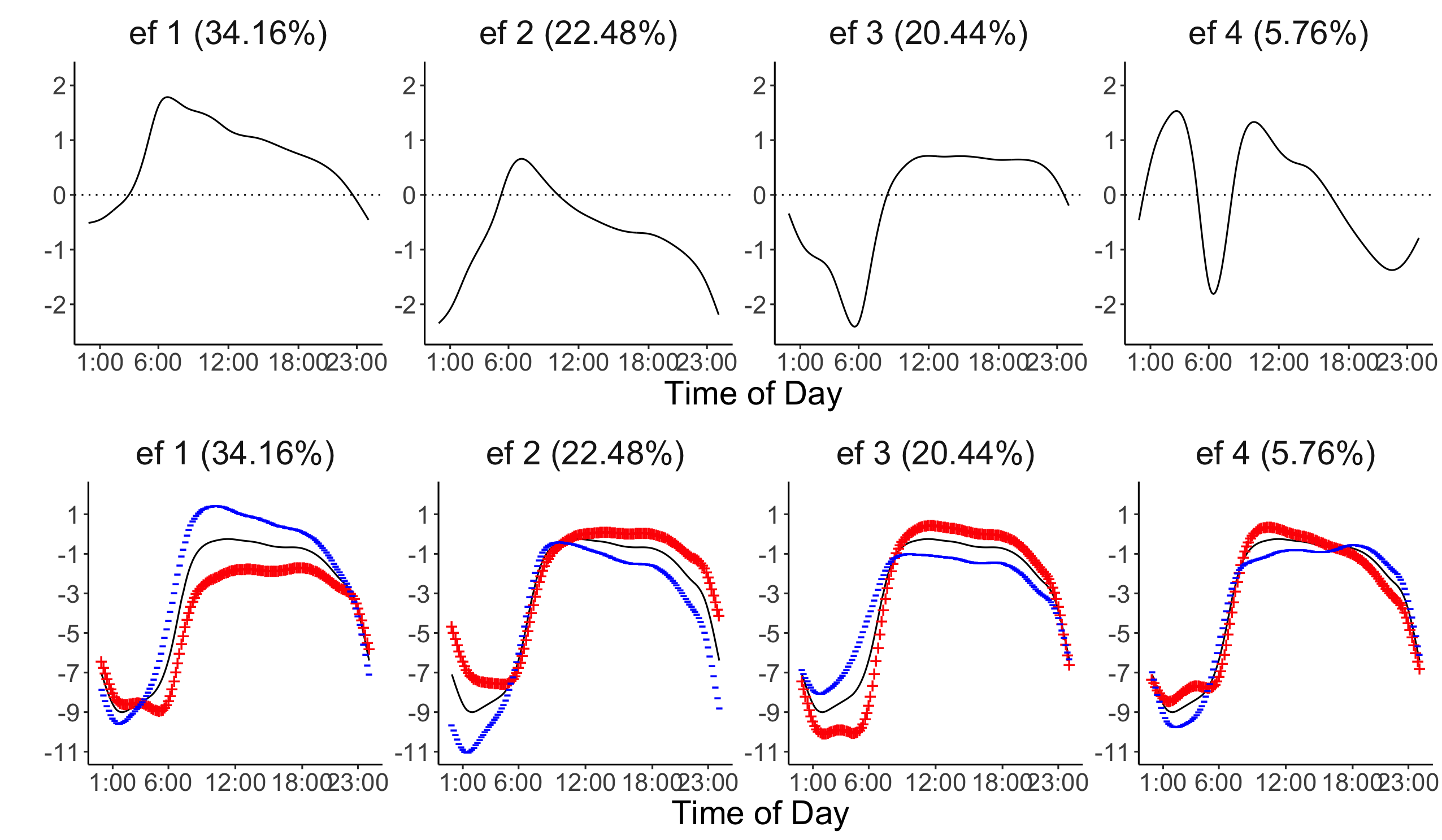}
  \caption{First row of panels: the first four estimated eigenfunctions from the NHANES dataset with the corresponding proportion of variability explained. Second row of panels:  population mean function for 60 years old males (black solid line) and the functions obtained by adding (red plus signs) and subtracting (blue minus signs) a one standard deviation multiple of the eigenfunctions.}
  \label{fig:nhanes-ef}
\end{figure}

The first row of panels in Figure~\ref{fig:nhanes-ef} displays the first four estimated eigenfunctions using GC-FPCA, which explain roughly $83$\% of the residual variability in the linear predictor after accounting for the fixed effects. To visualize the effect of eigenfunctions the second row of panels displays $\mu_{60}(s) \pm \sqrt{\widehat{\lambda}_l} \times \phi_l(s)$, where $\mu_{60}(s)$ is the mean of the linear predictors for $60$ year old males, and the $+$ and $-$ signs are shown in red and blue, respectively \citep{ramsaysilv2005,ramsayhook2009}.  

The first eigenfunction (first row, first column panel) is positive between $\sim$5:30AM and 10:30PM, which is typically associated with our our knowledge of human activity. Between roughly 10:30 PM and 5:30 AM, the eigenfunction is negative, corresponding to typical rest and sleep periods. This is not a surprising finding, but it confirms that active/inactive periods estimated from accelerometry provide reasonable results, at least in terms of the mean and first eigenfunction. The effect of a positive score on this eigenfunction on the mean (second row, first column panel) is to increase the likelihood of being active during the usual active hours and decrease the likelihood of being active during the usual sleeping periods. Recall that the eigenfunctions are only identified up to the sign and, in practice, one could obtain the symmetric plot of the eigenfunctions around the x-axis.

The second eigenfunction (first row, second panel) is positive only in the morning. The effect of a positive score on this eigenfunction on the mean (second row, second column panel) is to increase the likelihood of a person to be active in the morning, but decrease it during the night and the rest of the day. A potential example could be a $60$ year old male who has less disturbed sleep during the night, wakes up early in the morning, but is less active during the day than other $60$-year old males. This is intuitive, but identifying and quantifying such an effect from the data and ranking its variance explaining performance is a major step towards understanding the human active/inactive profiles in the complex, free-living environment. 

GC-FPCA was the only method that we could implement for model~\eqref{eq:NHANES_model} using the entire NHANES data set. On this data GC-FPCA took approximately $3$ minutes for Step 2 (models were parallelized), $11$ minutes for Step 3, and 5 days for Step 4. The computation time is dominated by fitting a generalized additive model (GAM) with four random effects (the scores on the eigenfunctions). This was implemented in the \texttt{mgcv::bam} function \citep{wood2017gam,wood2003,wood2004,wood2011}, which was not optimized for this model. Software used for this analysis is available in the supplementary materials. At this point, we have reasons to believe that efficient Bayesian MCMC may lead to further reductions in computation complexity and times by taking advantage of the conditional distributions of the a-priori independent random effects. Choosing one versus another method in Step 4 exceeds the scope of the current paper.  

As GFAMM implemented via the \texttt{refund::pffr} did not run on the entire NHANES data set, we ran it on a subset  of $200$ study participants. On this subset GFAMM took about one hour while GC-FPCA took about $3$ minutes. Both methods provide comparable estimates for the intercept. However, GFAMM fixed effects tend to be oversmoothed compared to GC-FPCA, even with the same number of basis functions. Just as in simulations, the computation time difference increases very fast with the number of study participants, but not with the number of observations per function.

\section{Discussion}\label{sec:discussion}
The NHANES 2011-2014 daily patterns of minute-level active/inactive analytic data set in this paper is a $8{,}700 \times {1,440}$ dimensional matrix with binary entries. Each row corresponds to a study participant, each column corresponds to a minute of the day (organized from midnight), and each entry indicates whether or not the person was active at that particular minute. These data are linked to demographic, behavioral, medical history and test results, as well as to mortality data (not used in this application) up to the latest release from the US Center of Disease Control (CDC). The problem addressed in this paper is to conduct statistical estimation and inference on the association between covariates and individual patterns of active/inactive patterns, while accounting for the sizeable, non-stationary, and complex within-person correlation of the data. This problem can be viewed as a function-on-scalar regression with non-Gaussian functional measurements (nomenclature introduced by \citep{reissfosr,reiss2017}; see also \citep{Wahba1983,ramsaysilv2005,crainiceanu2024book}). It can also be viewed as a Generalized Linear Mixed Model (GLMM) \citep{mcculloch2009,nlme} with a very large number of observations per trajectory. These data are released and are accompanied by fully reproducible vignette at (https://github.com/yulu98/GCFPCA).

This data structure and problem are neither isolated nor artificial. Consider, for example, the UK Biobank data that also collected high resolution wrist actigraphy data on close to $100{,}000$ study participants in the UK for up to seven days. Indeed, there are many other studies that now collect or transform data to high dimensional non-Gaussian vectors. Moreover, while the summaries provided by NHANES were continuous positive measurements (MIMS), one could argue that taking a threshold to obtain active/inactive profiles is an ``artificial way of creating a problem that did not exist." However, the thresholding is done for a very specific purpose, to make results interpretable and translatable into actionable information. As we have already mentioned in this paper, a recommendation of "100 MIMS units per day" lacks context, making it difficult to act upon, as the meaning of 100 MIMS units is unclear. 

We proposed the Generalized Conditional FPCA (GC-FPCA) to model such data. The method can be applied to the NHANES data set and the software can easily be modified to explore a variety of models. In contrast, the state-of-the-art GFAMM \citep{Scheipl2016GFADM} method did not run on the NHANES data set. In simulation studies on smaller data sets the statistical performance of GC-FPCA and GFAMM is comparable, while GC-FPCA is much faster. We wanted to emphasize that the main problem in these highly complex scenarios is not ``what method is best", but ``what method actually works and is implementable". GFAMM is an excellent method implemented in reproducible, transparent software. {\it GC-FPCA is another promising methodological and software tool in a landscape that has too few, not too many, such methods.}

\clearpage

\section*{Supplementary Material}
\setcounter{page}{1}
\setcounter{section}{0}
\setcounter{table}{0}
\setcounter{figure}{0}

\renewcommand{\thesection}{S.\arabic{section}}
\renewcommand{\thetable}{S.\arabic{table}}
\renewcommand{\theequation}{S.\arabic{equation}}
\renewcommand{\thefigure}{S.\arabic{figure}}

\section{Implementation}

\begin{description}

\item[R-package for GCFPCA:] The GCFPCA R package provides the code necessary to implement the GCFPCA model as described in our article. This package includes examples featured in the article, as well as a detailed vignette explaining usage and additional examples. The package website is available at \url{https://github.com/yulu98/GCFPCA}.

\end{description}

\section{Additional Simulation Results}

\subsection{Effect of bin width}

In this experiment, we examine the effect of varying the bin width while holding other parameters constant: sample size $I = 200$ and number of sampling points $K = 500$. For binary data, Fourier basis functions were used alongside binary fixed-effect covariates. Table \ref{supp-table:binary_bw} presents the simulation results for different bin widths, expressed as percentages of the total sampling points. As the bin width increased, computational time increased substantially. The MISE of the linear predictor initially decreased and then increased with larger bin widths, while the ISEs of the fixed-effect coefficient functions and the MISE of the principal components generally declined with increasing bin width. A bin width of approximately $10\%$ strikes a balance, maintaining fast computation times with higher accuracy in estimates.

\begin{table}[!htbp] \centering 
\footnotesize
\begin{tabular}{lcrrrrrr}
\toprule
bw (\%) & Time (min) & ${\rm MISE}(\eta)\times 10$ & ${\rm ISE}(\beta_0)\times 10^2$ & ${\rm AC}(\beta_0)$ & ${\rm ISE}(\beta_1)\times 10^2$ & ${\rm AC}(\beta_1)$ & ${\rm MISE}(\phi)\times 10$ \\
\midrule
2.5\%  & 3.00 & 6.54 & 2.16 & 0.94 & 4.29 & 0.94 & 0.37\\
5\%    & 4.98 & 6.00 & 2.08 & 0.95 & 4.20 & 0.94 & 0.29\\
10\%   & 9.96 & 5.70 & 2.06 & 0.95 & 4.15 & 0.95 & 0.24\\
15\%   & 12.28 & 5.71 & 2.06 & 0.95 & 4.16 & 0.95 & 0.22\\
20\%   & 14.65 & 5.83 & 2.06 & 0.95 & 4.16 & 0.95 & 0.22\\
\bottomrule
\end{tabular}
\caption{Simulation results for different bin width ($bw$), expressed as percentages of the whole functional domain, when $I=200{, }K=500$. The computation time (\textit{``Time(min)"}), MISE of $\eta$ (MISE($\eta$)), ISE of $\beta_{p}(s), p=0,1$ (``ISE($\beta_{r}$), $r=0,1$"), and AC of $\beta_{p}(s), p=0,1$ (``AC($\beta_{p}$), $p=0,1$"), MIS of $\phi$ (MISE($\phi$)). reported in the table, are median values across 1000 replications. ``-" stands for computing times that exceed 48 hours.}
\label{supp-table:binary_bw}
\end{table}

\subsection{Type of eigenfunctions}

\begin{figure}[!ht]
\centering
\includegraphics[width=\textwidth]{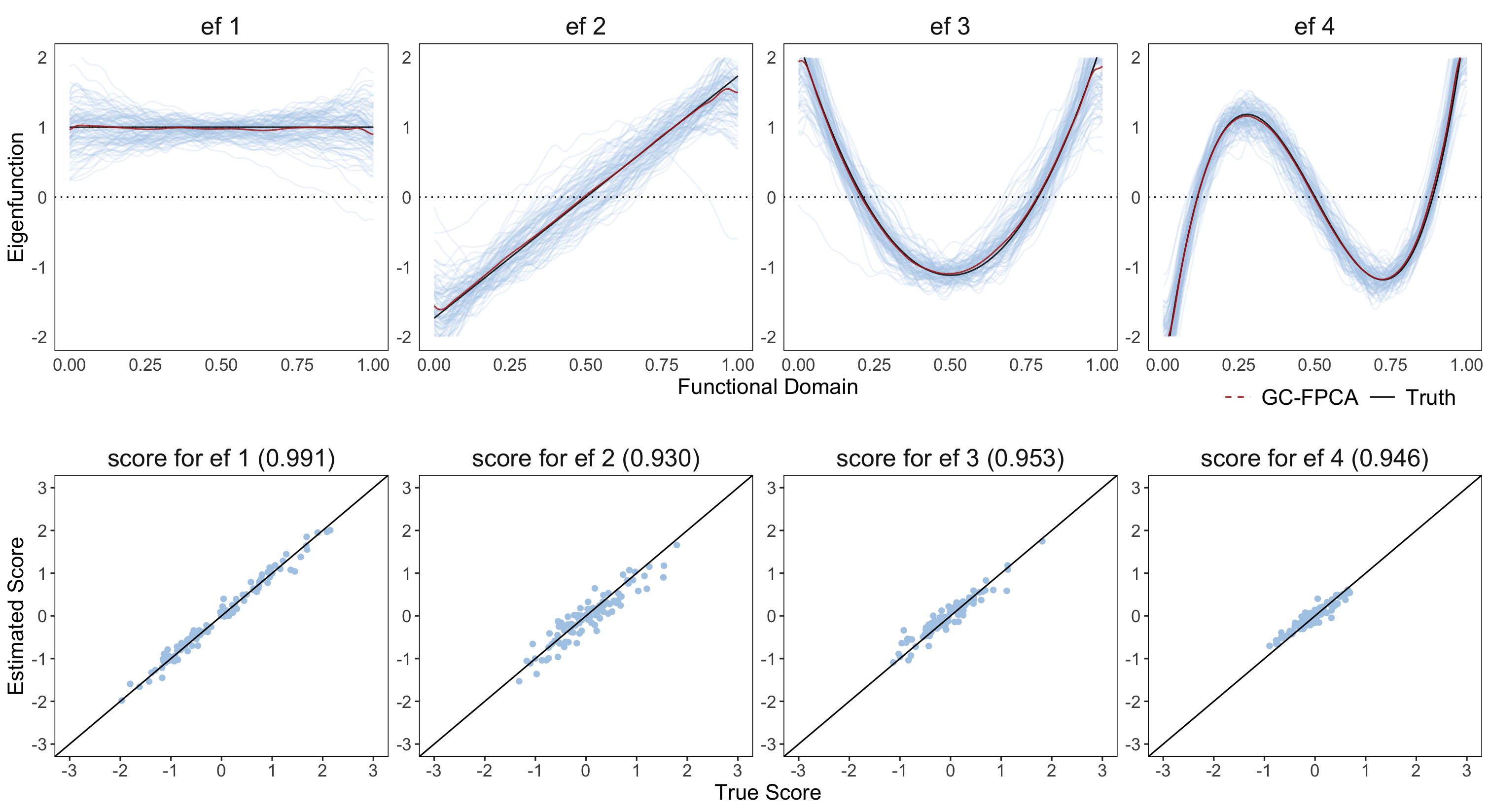}
  \caption{Simulation results of 1000 simulations when $ I = 100, K = 1000$ (binary functional data using orthogonal polynomials (non-periodic) eigenfunctions). First row of panels: $100$ estimated eigenfunctions (red), true eigenfunction (black solid lines), average over $1{,}000$ GC-FPCA estimates of the eigenfunctions (dashed, red line). Second row of panels: scatter plots of true (x-axis) versus estimated (y-axis) scores (blue dots) with their correlation (in the parentheses), identity line (black solid line) for one simulated data set.} 
  \label{supp-fig:ef_np}
\end{figure}

This experiment investigates the impact of using different eigenfunction bases, namely orthogonal polynomials (non-periodic) and Fourier basis (periodic). Here, the sample size is fixed at $I = 1000$, with $K = 100$ sampling points and a bin width of $10\%$ of the total sampling points. Binary data were generated with binary fixed-effect covariates. Table \ref{supp-table:(non)periodic_simulation} displays simulation outcomes for each eigenfunction type, while Figure \ref{supp-fig:ef_np} illustrates eigenfunction fits and score correlations using the orthogonal polynomial basis (non-periodic).

\begin{table}[!htbp] \centering 
\footnotesize
\begin{tabular}{lcrrrrrr}
\toprule
Ef type & Time (min) & ${\rm MISE}(\eta)\times 10$  & ${\rm ISE}(\beta_0)\times 10^2$ & ${\rm AC}(\beta_0)$ & ${\rm ISE}(\beta_1)\times 10^2$ & ${\rm AC}(\beta_1)$ & ${\rm MISE}(\phi)\times 10$\\
\midrule

Periodic & 0.13 & 0.18 & 3.85 & 0.94 & 7.81 & 0.94 & 0.44\\
\midrule
Non-periodic & 3.75 & 0.40 & 4.10 & 0.79 & 8.40& 0.94 & 0.49\\
\bottomrule
\end{tabular}
\caption{Simulation results for different types of eigenfunctions when $I = 1000, \text{, }K=100$. The computation time (\textit{``Time(min
)"}), MISE of $\eta$ (MISE($\eta$)), ISE of $\beta_{p}(s), p=0,1$ (``ISE($\beta_{r}$), $r=0,1$"), and AC of $\beta_{p}(s), p=0,1$ (``AC($\beta_{p}$), $p=0,1$"), MISE of $\phi$ (MISE($\phi$))reported in the table, are median values across 1000 replications.}

\label{supp-table:(non)periodic_simulation}
\end{table}

\subsection{Type of Fixed effect covariates}

This experiment evaluates the impact of fixed-effect covariates generated from $\rm{Bernoulli}(\frac{1}{2})$ and $N(0,1)$ distributions. With sample size fixed at $I = 100$, $K = 100$ sampling points, and bin width set at $5\%$ of total sampling points, Fourier basis functions and binary outcomes are used. Table \ref{supp-table:fixed_effect} reports the results. Continuous covariates yield lower ISEs for fixed-effect functions, while the linear predictor and eigenfunction estimates remain similar to the binary case.

\begin{table}[!htbp] \centering 
\footnotesize
\begin{tabular}{lcrrrrrr}
\toprule

X type & Time (min) & ${\rm MISE}(\eta)\times 10$  & ${\rm ISE}(\beta_0)\times 10^2$ & ${\rm AC}(\beta_0)$ & ${\rm ISE}(\beta_1)\times 10^2$ & ${\rm AC}(\beta_1)$ & ${\rm MISE}(\phi)\times 10$\\
\midrule
Binary & 0.06 & 3.31 & 5.79 & 0.91 & 11.48 & 0.91 & 2.04 \\
\midrule
Continuous & 0.21 & 3.47 & 3.30 & 0.91 & 3.49 & 0.92 & 2.16 \\

\bottomrule
\end{tabular}
\caption{Simulation results for different types of eigenfunctions when $I = 100, \text{, }K=100$. The computation time (\textit{``Time(min
)"}), MISE of $\eta$ (MISE($\eta$)), ISE of $\beta_{p}(s), p=0,1$ (``ISE($\beta_{r}$), $r=0,1$"), and AC of $\beta_{p}(s), p=0,1$ (``AC($\beta_{p}$), $p=0,1$"), MISE of $\phi$ (MISE($\phi$))reported in the table, are median values across 1000 replications.}
\label{supp-table:fixed_effect}
\end{table}

\subsection{Gaussian Responses}

In this experiment, Gaussian-distributed outcomes are considered. Table \ref{supp-table:gaussian_accuracy_I} shows results for varying sample sizes, $I$, with fixed $K = 100$ observations per function. Table \ref{supp-table:gaussian_accuracy_K} details outcomes for varying $K$, holding $I = 100$ constant. Table \ref{supp-table:gaussian_pc} presents eigenfunction estimates across different $I$ and $K$ combinations. The performance metrics are consistent with those used for binary outcomes.

\begin{table}[!htbp] \centering 
\footnotesize
\begin{tabular}{lcrrrrrr}
\toprule
I & Method & Time (min) & ${\rm MISE}(\eta)\times 10$  & ${\rm ISE}(\beta_0)\times 10^2$ & ${\rm AC}(\beta_0)$ & ${\rm ISE}(\beta_1)\times 10^2$ & ${\rm AC}(\beta_1)$ \\
\midrule
\multirow{2}{*}{100} & GC-FPCA & 0.08 & 0.54 & 1.16 & 0.93 & 2.23 & 0.93 \\
& GFAMM & 23.37 & 0.78 & 4.22 & 0.91 & 8.45 & 0.93 \\
\midrule
\multirow{2}{*}{200} & GC-FPCA & 0.15 & 0.47 & 0.58 & 0.93 & 1.12 & 0.93 \\
& GFAMM & 205.46 & 0.77 & 2.06 & 0.92 & 4.15 & 0.93 \\
\midrule
\multirow{2}{*}{500} & GC-FPCA & 1.17 & 0.43 & 0.25 & 0.93 & 0.49 & 0.93 \\
& GFAMM & - & - & - & - & - & -  \\
\midrule
\multirow{2}{*}{1000} & GC-FPCA & 8.18 & 0.41 & 0.15 & 0.92 & 0.26 & 0.93  \\
& GFAMM & - & - & - & - & - & - \\
\bottomrule
\end{tabular}
\caption{Simulation results for different $ I $ when $ K=100 $. The computation time (\textit{``Time(min)"}) and performance metrics like MISE of $\eta$, ISE and AC for $\beta_{0}$ and $\beta_{1}$, and MISE of $\phi$ are reported as median values across 1000 replications. ``-" indicates computing times exceeding 48 hours.}
\label{supp-table:gaussian_accuracy_I}
\end{table}

\begin{table}[!htbp] \centering 
\footnotesize
\begin{tabular}{lcrrrrrr}
\toprule
K & Method & Time (min) & $MISE(\eta)\times 10$  & $ISE(\beta_0)\times 10^2$ & $AC(\beta_0)$ & $ISE(\beta_1)\times 10^2$ & $AC(\beta_1)$ \\
\midrule
\multirow{2}{*}{200} & GC-FPCA & 0.17 & 0.25 & 0.94 & 0.93 & 1.84 & 0.93  \\
& GFAMM & 24.60 & 4.32 & 4.00 & 0.91 & 8.06 & 0.93 \\
\midrule
\multirow{2}{*}{500} & GC-FPCA & 0.37 & 0.10 & 0.79 & 0.94 & 1.57 & 0.94 \\
& GFAMM & 26.53 & 1.99 & 3.86 & 0.90 & 7.82 & 0.93\\
\midrule
\multirow{2}{*}{1000} & GC-FPCA & 0.97 & 0.05 & 0.75 & 0.94 & 1.49 & 0.93 \\
& GFAMM & 27.18 & 1.12 & 3.41 & 0.91 & 8.12 & 0.93  \\
\bottomrule
\end{tabular}
\caption{Simulation results for different $ K $ when $ I=100 $. The computation time (`Time(min)`), MISE of $\eta$ (MISE($\eta$)), ISE of $\beta_{r}(s), r=0,1$ (ISE($\beta_{p}$), $p=0,1$), and AC of $\beta_{p}(s), p=0,1$ (AC($\beta_{p}$), $p=0,1$), reported in the table, are median values across 1000 replications.}
\label{supp-table:gaussian_accuracy_K}
\end{table}

\begin{table}[!htbp]
\begin{center}
\begin{tabular}{ccc|ccc}
\toprule
I & K & ${\rm MISE}(\phi) \times 10$ & I & K &  ${\rm MISE}(\phi)\times 10$\\ \midrule
 100 & 100& 3.59 &   &   &   \\
 200&100& 2.57 & 100&200 & 1.53 \\
 500&100& 1.98 & 100&500 & 0.53\\
 1000&100& 1.80 & 100&1000 & 0.26\\

\bottomrule
\end{tabular}
\end{center}
\caption{Median MISE ($\times 10$) across $1{,}000$ simulations  for estimating the eigenfunctions for different $I$ and $K$ (binary functional data using Fourier eigenfunctions).}
\label{supp-table:gaussian_pc}
\end{table}

\subsection{Poisson Responses}

In this experiment, we generated Poisson-distributed outcomes. Table \ref{supp-table:poisson_accuracy_I} shows results for varying sample sizes, $I$, with fixed $K = 100$ observations per function. Table \ref{supp-table:poisson_accuracy_K} details outcomes for varying $K$, holding $I = 100$ constant. Table \ref{supp-table:poisson_pc} presents eigenfunction estimates across different $I$ and $K$ combinations. The performance metrics are consistent with those used for binary outcomes.

\begin{table}[!htbp] \centering 
\footnotesize
\begin{tabular}{lcrrrrrr}
\toprule

I & Method & Time (min) & ${\rm MISE}(\eta)\times 10$  & ${\rm ISE}(\beta_0)\times 10^2$ & ${\rm AC}(\beta_0)$ & ${\rm ISE}(\beta_1)\times 10^2$ & ${\rm AC}(\beta_1)$\\
\midrule
\multirow{2}{*}{100} & GC-FPCA & 0.21 & 0.66 & 5.02 & 0.90 & 8.68 & 0.92 \\
& GFAMM & 46.17 & 1.17 & 4.70 & 0.93 & 8.08 & 0.93 \\
\midrule
\multirow{2}{*}{200} & GC-FPCA & 0.50 & 0.57 & 2.79 & 0.93 & 4.16 & 0.93 \\
 & GFAMM & 345.67 & 1.15 & 2.45 & 0.92 & 3.92 & 0.93 \\
\midrule
\multirow{2}{*}{500} & GC-FPCA & 2.39 & 0.50 & 1.49 & 0.93 & 1.67 & 0.94 \\
 & GFAMM & - & - & - & - & - & - \\
\midrule
\multirow{2}{*}{1000} & GC-FPCA & 12.98 & 0.48 & 1.13 & 0.92 & 0.89 & 0.93 \\
 & GFAMM & - & - & - & - & - & - \\
\bottomrule
\end{tabular}
\caption{Simulation results for different $ I $ when $ K=100 $. The computation time (\textit{``Time(min)"}), MISE of $\eta$ (MISE($\eta$)), ISE of $\beta_{p}(s), p=0,1$ (``ISE($\beta_{r}$), $r=0,1$"), and AC of $\beta_{p}(s), p=0,1$ (``AC($\beta_{p}$), $p=0,1$"), reported in the table, are median values across 1000 replications. ``-" stands for computing times exceed 48 hours.}

\label{supp-table:poisson_accuracy_I}
\end{table}

\begin{table}[!htbp] \centering 
\footnotesize
\begin{tabular}{lcrrrrrr}\toprule

K & Method & Time (min) & $MISE(\eta)\times 10$  & $ISE(\beta_0)\times 10^2$ & $AC(\beta_0)$ & $ISE(\beta_1)\times 10^2$ & $AC(\beta_1)$\\ 
\midrule

\multirow{2}{*}{200} & GC-FPCA & 0.45 & 0.33 & 4.36 & 0.92 & 8.48 & 0.93 \\

& GFAMM & 38.64 & 0.699 & 4.19 & 0.89 & 7.84 & 0.93  \\
\midrule

\multirow{2}{*}{500} & GC-FPCA & 1.56 & 0.15 & 4.44 & 0.91 & 9.40 & 0.91 \\

& GFAMM & 39.32 & 0.341 & 3.87 & 0.89 & 7.68 & 0.93 \\
\midrule

\multirow{2}{*}{1000} & GC-FPCA & 5.14 & 0.08 & 4.45 & 0.92 & 9.60 & 0.91 \\

& GFAMM & 41.63 & 0.195 & 3.80 & 0.89 & 7.65 & 0.92 \\
\bottomrule
\end{tabular}
\caption{Simulation results for different $ K $ when $ I=100 $. The computation time (``Time(min)"), MISE of $\eta$ (``MISE($\eta$)"), ISE of $\beta_{r}(s), r=0,1$ (``ISE($\beta_{p}$), $p=0,1$"), and AC of $\beta_{p}(s)$, $p=0,1$"), reported in the table, are median values across 1000 replications.}
\label{supp-table:poisson_accuracy_K}
\end{table}

\begin{table}[ht]
\begin{center}

\begin{tabular}{ccc|ccc}
\toprule
I & K & ${\rm MISE}(\phi) \times 10$ & I & K &  ${\rm MISE}(\phi)\times 10$\\ \midrule
 100 & 100& 0.80 &   &   &   \\
 200&100& 0.48 & 100&200&0.59\\
 500&100& 0.29 & 100&500&0.49\\
 1000&100& 0.23 & 100&1000 &0.44\\

\bottomrule
\end{tabular}

\end{center}
\caption{Median MISE ($\times 10$) across $1{,}000$ simulations  for estimating the eigenfunctions for different $I$ and $K$ (binary functional data using Fourier eigenfunctions).}
\label{supp-table:poisson_pc}
\end{table}

\subsection{Additional Simulation results}

\begin{figure}[!ht]
\centering
\includegraphics[width=\textwidth]{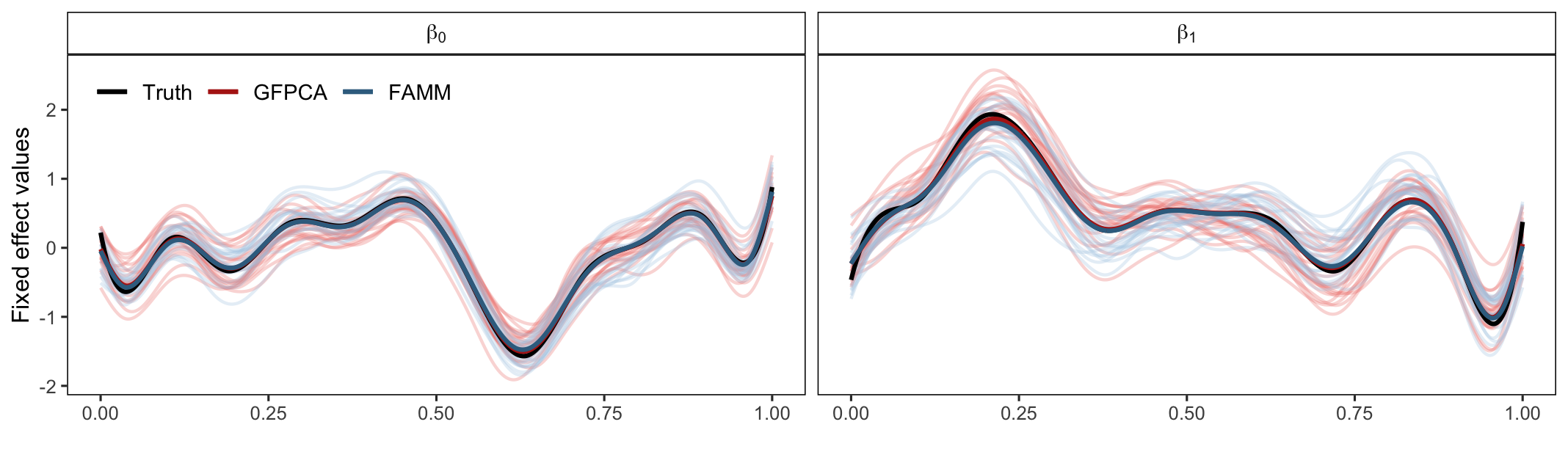}
  \caption{Comparison of simulation results between GC-FPCA (red) and GFAMM (blue) for binary functional data generated using Fourier eigenfunctions. The plot shows 20 estimated fixed-effect functions (light transparent colored lines), the true effect functions (solid black lines), and the average fixed-effect functions over 1,000 GC-FPCA/GFAMM estimates (solid colored lines).} 
  \label{supp-fig:fe_fit_simu}
\end{figure}

Figure \ref{supp-fig:fe_fit_simu} display 20 estimated fixed-effect coefficient functions for both GC-FPCA and GFAMM models, along with the average fixed-effect coefficient over 1,000 simulations. These estimates are compared to the true generating functions to evaluate the models' accuracy.

\section{More NHANES data results}

\subsection{Data Summary}

Here we show a summary of demographic variables and traditional risk factors for the data we used.

{\footnotesize
\begin{longtable}{lc} 
\label{supp-table:table1}
\\ \toprule\toprule
 & Mean (SD) or N (\%)\\\midrule
Sample Size & 8700 \\\midrule
Age   &       49.47 (17.51) \\
Gender & \\
\;\;Female &         4529 (52.1) \\
BMI & \\                                 
\;\;Normal    &  2513 (28.9) \\
\;\;Underweight   &   164 ( 1.9) \\
\;\;Overweight    &  2780 (32.0) \\
\;\;Obese         &  3243 (37.3) \\
Race & \\                                
\;\;Non-Hispanic White   &     3544 (40.7) \\
\;\;Non-Hispanic Black   &     2058 (23.7) \\
\;\;Mexican American     &     1013 (11.6) \\
\;\;Non-Hispanic Asian   &     1010 (11.6) \\
\;\;Other Hispanic       &      826 ( 9.5) \\
\;\;Other Race           &      249 ( 2.9) \\
Overall Health           & \\   
\;\;Excellent            &     3516 (40.4) \\
\;\;Very good            &      849 ( 9.8) \\
\;\;Good                 &     2324 (26.7) \\
\;\;Fair                 &     1703 (19.6) \\
\;\;Poor                 &      308 ( 3.5) \\
Diabetes & \\                            
\;\;No                   &     7325 (84.2) \\
\;\;Borderline           &      237 ( 2.7) \\
\;\;Yes                  &     1138 (13.1) \\
Arthritis & \\                           
\;\;Yes                  &     2317 (26.6) \\
CHF & \\                                 
\;\;Yes                  &      284 ( 3.3) \\
CHD & \\                                 
\;\;Yes                  &      349 ( 4.0) \\
Heart Attack & \\                        
\;\;Yes                  &      324 ( 3.7) \\
Stroke & \\                              
\;\;Yes                  &      317 ( 3.6) \\
Cancer & \\                              
\;\;Yes                  &      830 ( 9.5) \\
Alcohol Consumption & \\             
\;\;Never drinker       &     1207 (13.9) \\
\;\;Former drinker      &     1435 (16.5) \\
\;\;Moderate drinker    &     4865 (55.9) \\
\;\;Heavy drinker       &      559 ( 6.4) \\
\;\;Not Available       &      634 ( 7.3) \\
Smoking & \\                   
\;\;Never               &     4945 (56.8) \\
\;\;Former              &     2031 (23.3) \\
\;\;Current             &     1724 (19.8) \\
Mobility Problem & \\                    
\;\;No difficulty      &     6995 (80.4) \\
\;\;Any difficulty      &     1705 (19.6) \\
\bottomrule\bottomrule
\\ \caption{Summary of demographic variables and traditional risk factors for subjects from the NHANES 2011-2014 accelerometry dataset.}
\end{longtable}}

\subsection{Scree plot for selecting number of eigenfunctions}

\begin{figure}[!tbh]
\centering
\includegraphics[width=\textwidth]{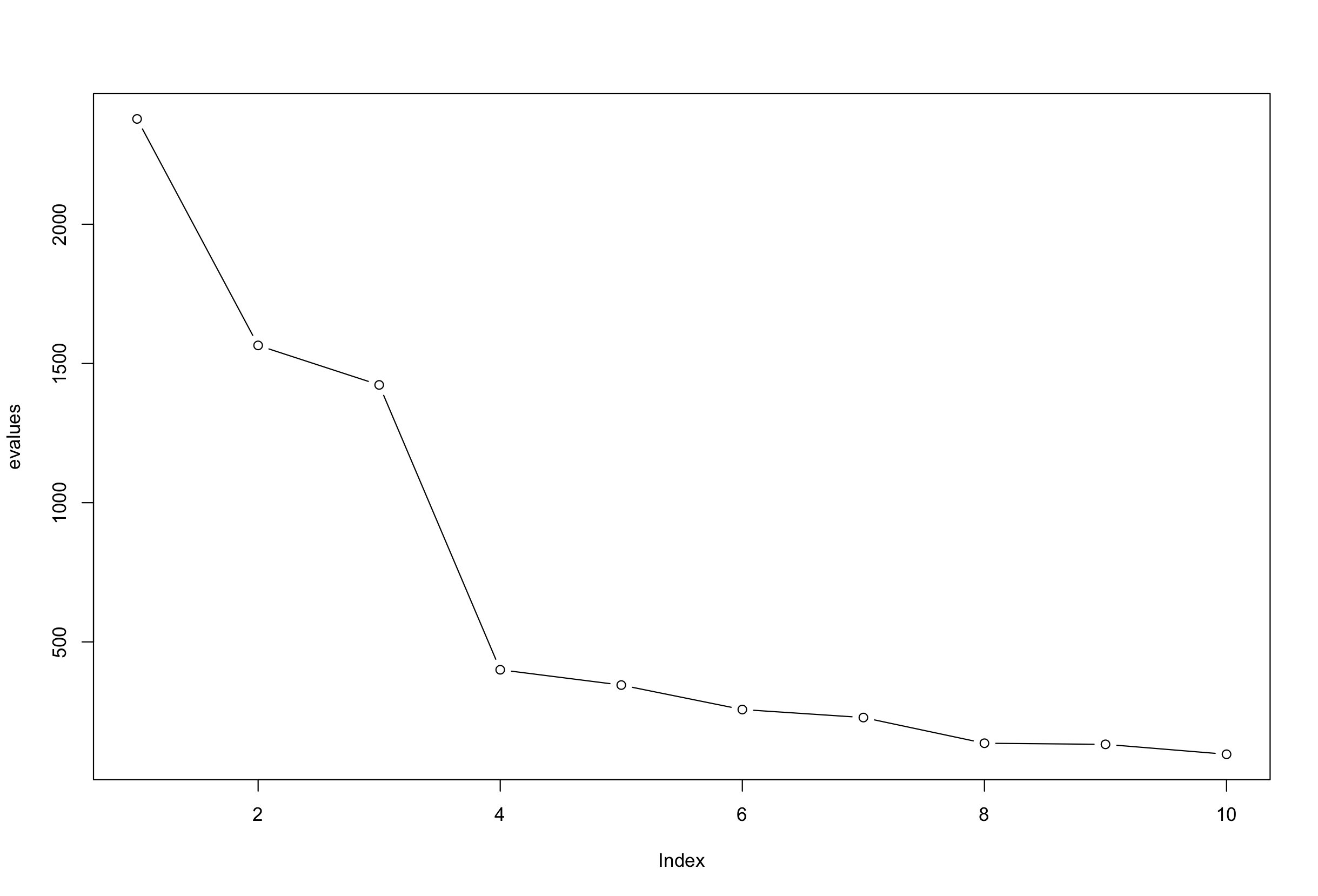}
  \caption{Scree plot of the eigenvalues of the first 10 principal components.} 
  \label{supp-fig:scree}
\end{figure}

Figure \ref{supp-fig:scree} displays the eigenvalues of the first $10$ principal components. An "elbow" appears at the $4$-th principal component, indicating a point of diminishing returns. Based on this, we selected $4$ eigenfunctions for the GC-FPCA model.

\subsection{Comparison with GFAMM on subsampled data}

\begin{figure}[!tbh]
\centering
\includegraphics[width=\textwidth]{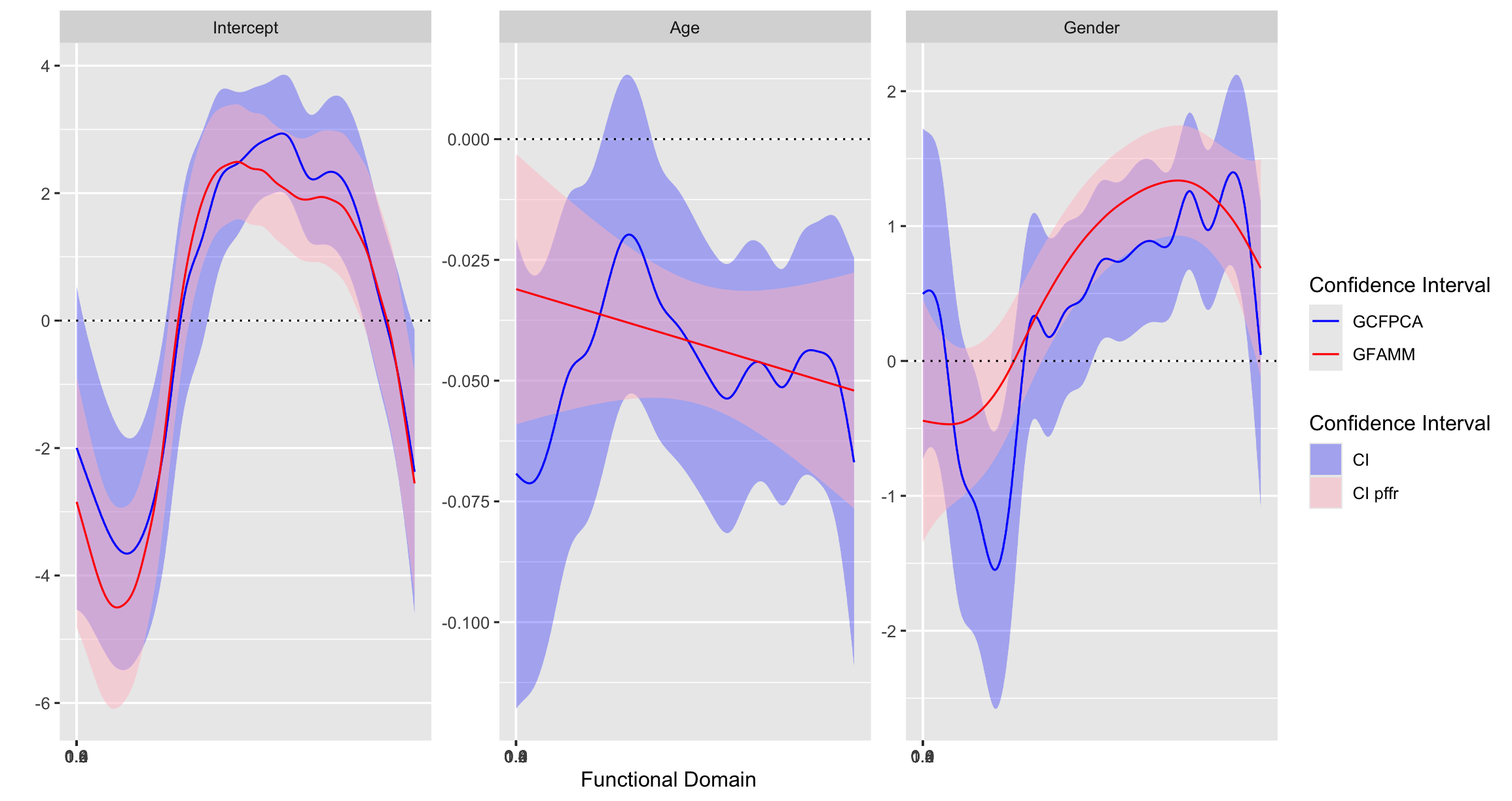}
  \caption{The estimated fixed effects obtained using GC-FPCA (blue solid lines) on a subsample of NHANES data are shown alongside their $95\%$ pointwise confidence intervals, represented by blue shading. GFAMM-derived fixed effect estimates are displayed as red solid lines, with corresponding $95\%$ confidence intervals indicated by red shading. The panels, from left to right, present estimates for the intercept, baseline age, and gender (female).} 
  \label{supp-fig:fe_fit}
\end{figure}

We selected a subsample of 200 subjects from the NHANES population to compare fixed effect estimates using the GC-FPCA and GFAMM methods. For GFAMM, we utilized 30 cubic regression basis functions for both fixed and random effects. Figure~\ref{supp-fig:fe_fit} illustrates the estimated coefficients for each method along with $95\%$ pointwise confidence intervals (shaded areas). GC-FPCA estimates are shown with blue lines and shaded areas, while GFAMM estimates are represented in red. Both methods provide comparable estimates for the intercept. However, GFAMM fixed effects tend to be oversmoothed compared to GC-FPCA, even with the same number of basis functions.


\bibliographystyle{biom}
\bibliography{ref}

\end{document}